\documentclass[11pt]{article}
\usepackage{graphicx}
\usepackage{cancel}[]
\usepackage{amssymb,amsmath}
\usepackage{ifpdf}
\usepackage{subfigure}
\usepackage{hyperref}[]
\usepackage{fullpage}
\usepackage{color}
\usepackage{ulem}
\ifpdf
    \DeclareGraphicsExtensions{.pdf}
\else
    \DeclareGraphicsExtensions{.eps}
\fi

\begin{document}

\vspace{0.1 in}
\centerline{\Large{\bf{Use of  Slow Light to test the Isotropy of Space}}}
\vspace{0.1 in}
\centerline{\rm{ A. C. Melissinos}}

\centerline{\today}

\vspace{0.1 in}

Slow light refers to light signals that propagate through particular media
with low group velocity \cite{Boyd}. We propose to exploit this low velocity
as compared to the motion of the Earth, to test for the isotropic propagation
of light through space. The motion of the Earth consists of three components:
(a) due to the Earth's rotation on it's axis (at the latitude of Rochester NY)
$\beta_1 = v_1/c = 1.13\times
10^{-6}$, (b) due to the Earth's orbital motion around the sun, $\beta_2 =
v_2/c = 1\times 10^{-4}$, and (c) due to the motion of the solar system
with respect to the preferred microwave background frame, $\beta_3
 = v_3/c = 1.3\times 10^{-3}$. It is also known that a moving medium
 will ``drag" the light signals passing through it in both the longitudinal
 and transverse direction \cite{Fizeau, Jones}. In fact, the latter case has been
 recently demonstrated using slow light \cite{Padgett}, and has prompted the
 present proposal.\\

 Consider a Michelson or Mach-Zehnder interferometer with a slow medium
 of length $L$ in one arm and group index $n_g$ \cite{Shi}. If the
 interferometer is moving with velocity $v_m =c\beta_m$, along the direction
 of the arm containing the slow medium, the group
 velocity of the light in the medium will be modified  due to ``aether drag",
 \begin{equation}
 \beta_g = \frac{1/n_g + \beta_m}{1 + \beta_m/n_g}
 \end{equation}
 For an ordinary medium of refractive index $n$, the light traversing the
 medium in the arm, acquires a phase
 \begin{equation}
 \Delta \phi = \frac{\omega}{c} n L
 \end{equation}
 where $\omega = 2\pi \nu$ is the angular frequency of the light. If the
 medium has a large group index, the phase  change due to a change in
 frequency is given by \cite{Shi}
 \begin{equation}
 \frac{d\Delta \phi}{d\omega} = \frac{d}{d\omega}(\frac{\omega n L}{c})
 = \frac{L}{c}( n + \omega\frac{dn}{d\omega})= \frac{Ln_g}{c}
 \end{equation}
 In a similar way we can find the change in phase due to a change in the
 group velocity. Since
 \begin{equation}
 \Delta \phi = \omega\ t = \omega \frac{L}{v_g}
 \end{equation}
 \begin{equation}
 \frac{d\Delta \phi}{d v_g} = -\omega\frac{L}{v^2_g} \qquad {\rm{or}}
 \qquad \frac{1}{2\pi}\Delta \phi = - \frac{L}{\lambda}\frac{1}{\beta_g}
 \frac{dv_g}{v_g}
 \end{equation}\\

 We assume that a group index $ n_g = 10^3$ can be achieved. If the
 slow light arm is in the E-W direction, and considering only $\beta_1$
 and $\beta_2$ the effective velocity of the slow medium is
 \begin{equation}
 \beta_g \simeq 1/n_g + \beta_m \simeq 1/n_g + \beta_2 + \beta_1 {\rm{cos}}
 (\Omega_s t)
 \end{equation}
 The effective velocity $\beta_g$ is dominated by $1/n_g$ but it is modulated
 at the sidereal frequency $\nu_s =\Omega_s/2\pi$ with an amplitude
 of 340 m/s. Returning to Eq.(5) and setting $L=0.3$ m, $\lambda = 10^{-6}$
 m, and if the phase change $\Delta \phi/2 \pi$ is measured to one
 radian, the resolution in $v_g$ is
 \begin{equation}
 dv_g = \frac{\lambda}{L} \beta_g v_g = 10^{-3}\ \rm{m/s}
 \end{equation}
 where we used $\beta_g = 10^{-3}$, namely the velocity of the slow light,
 and not the much smaller velocity due to the ``aether" drag. Thus the
 relative resolution in measuring the velocity of the Earth's rotation
 exceeds one part in $10^{5}$. Of course the phase shift can be measured 
 to much better accuracy than assumed so far.\\

 As indicated in Eq.(6) the phase shift at the sidereal frequency
 is only $1/10^3$ of the overall shift induced by the presence of
 the slow medium. This ``static" shift is set to null
 by adjusting the interferometer. The null condition must be
 maintained during the duration of the measurement, and it's fluctuations
 could mask the signal. However, since we are looking for a narrow
 line in the frequency domain, it would be easy to extract the
 signal from the noise, given sufficient integration time; several days
 of data are needed to identify the sidereal frequency.
 To maintain the interferometer on the dark fringe it will be necessary
 to feedback on the arm length. Various schemes can be considered,
 but the simplest may be to lock slightly off the dark fringe
 so that the interferometer output includes a signature of the
 detuning. Given the very low frequency of interest, high frequency
 noise can be strongly filtered.\\

 The most recent results on the isotropy of space, obtained by a
 Michelson type experiment \cite{Hall} set a limit of $v/c <
4\times 10^{-8}$ or $v \leq 10 $ m/s. The surprising improvement
using slow light is because in this case the measured
effect depends to first order in $v$ rather than on $(v/c)^2$,
and the slow modulation is provided by the Earth's rotation,
that is well known from astronomical data. A deviation from the
expected waveform would be an indication of anisotropic propagation
of light.\\

\end{document}